\documentclass[fleqn,usenatbib]{mnras}

\usepackage{newtxtext,newtxmath}

\usepackage[T1]{fontenc}

\DeclareRobustCommand{\VAN}[3]{#2}
\let\VANthebibliography\thebibliography
\def\thebibliography{\DeclareRobustCommand{\VAN}[3]{##3}\VANthebibliography}
\defcitealias{pynbody}{Pynbody}

\usepackage{graphicx}	
\usepackage{amsmath}	
\usepackage{caption}
\usepackage{xcolor}
\usepackage{subcaption}
\usepackage[normalem]{ulem} 

\newcommand{\soutPC}{\bgroup\markoverwith{\textcolor{cyan}
{\rule[0.5ex]{2pt}{1pt}}}\ULon}

\newcommand{\soutdif}{\bgroup\markoverwith{\textcolor{magenta}{\rule[0.5ex]{2pt}{1pt}}}\ULon}

\title[The fate of primordial discs]{Exploring the fate of primordial discs in Milky Way-sized galaxies with the GigaEris simulation}

\author[F. van Donkelaar et al.] 
{Floor van Donkelaar,$^{1}$\thanks{floor.vandonkelaar@uzh.ch} Lucio Mayer,$^{1}$ Pedro R. Capelo$^{1}$  and Piero Madau$^{2,3}$\\
$^1$Department of Astrophysics, University of Zurich, Winterthurerstrasse 190, CH-8057 Z\"urich, Switzerland\\
$^2$Department of Astronomy and Astrophysics, University of California, 1156 High Street, Santa Cruz, CA 95064, USA\\
$^3$Dipartimento di Fisica ``G. Occhialini", Universit\`a degli Studi di Milano-Bicocca, P.za della Scienza 3, I-20126 Milano, Italy}

\date{Accepted XXX. Received YYY; in original form ZZZ}

\pubyear{2024}

\begin{document}
\label{firstpage}
\pagerange{\pageref{firstpage}--\pageref{lastpage}}
\maketitle

\begin{abstract}
Recent observations with JWST and ALMA have unveiled galaxies with regular discs at significantly higher redshifts than previously expected. This appears to be in contrast with constraints on the stellar populations of the Milky Way, suggesting that the bulk of the Galactic thin disc formed after $z=1$, and raises questions about the history, evolution, and survivability of primordial discs. Here, we use GigaEris, a state-of-the-art $N$-body, hydrodynamical, cosmological ``zoom-in'' simulation with a billion particles within the virial radius, to delve into the formation of the early kinematically cold discs (KCDs), defined by their ratio between the mean rotational velocity and the radial velocity dispersion, of a Milky Way-sized galaxy at redshifts $z\gtrsim 4$. Our analysis reveals a primarily inward migration pattern for disc stars formed at $z \gtrsim 6$, turning into a mix of inward and outward migration at later times. Stars migrating outwards undergo minimal kinematic heating, and might be identified as part of the thin disc forming at much later epochs. We find that approximately 76 per cent of all stars formed in the KCD at $z \sim 7$ become part of a pseudo-bulge by $z = 4.4$. This proportion decreases to below 10 per cent for KCD stars formed at $z \lesssim 5$. The inward migration of stars born in our KCDs at $z \gtrsim 4$ deviates from the expected inside-out formation scenario of thin discs at lower redshifts. Our results suggest a novel ``two-phase'' disc formation process, whereby the early disc transforms primarily into the pseudo-bulge within less than a billion years, whereas the present-day disc forms subsequently from higher-angular momentum material accreted at later times.
\end{abstract}

\begin{keywords}
Galaxy: evolution - Galaxy: disc - Galaxy: Bulge - galaxies: high-redshift - methods: numerical

\end{keywords}



\section{Introduction}

Since the initial proposal by \citet{Gilmore:1983aa} of a dual-component structure within the disc of the Milky Way (MW), numerous observational studies have supported the idea of distinct thin- and thick-disc components. Thin-disc stars, located closer to the plane and with low-eccentricity orbits, exhibit low [$\alpha$/Fe] enhancement and a metallicity distribution centred around solar amounts and ranging from supersolar values down to [Fe/H] $\sim -0.7$  \citep[e.g.][]{Robin:1996aa, Norris:1999aa, ojha:2001aa, Larsen:2003aa, juric:2008aa, Bensby:2014aa, Recio:2014aa, Hayden:2015aa}. In contrast, the thick disc extends to larger distances from the plane and its stars exhibit higher [$\alpha$/Fe] enhancements and a lower metallicity content than those of the thin disc, with a broader metallicity distribution centred around [Fe/H] $\sim -0.55$  \citep[e.g.][]{Gilmore:1989aa, katz:2011aa}. Kinematically, thick-disc stars have larger velocity dispersions than the stars in the thin disc \citep[e.g.][]{ Gilmore:2002aa, Soubiran:2003aa, Parker:2004aa, Wyse:2006aa, Mackereth:2019aa}. Despite these differences, there is considerable overlap between the two components, and the origin of the thick and thin discs remain the subject of ongoing debate \citep[see, e.g.][]{Quinn:1993aa,Abadi:2003aa, Brook:2005aa, Villalobos:2008aa, Schonrich:2009aa, Schonrich:2009ab, Aumer:2016aa, Clarke:2019aa, Debattista:2019aa, Buck:2020aa, Agertz:2021aa}. 

Yet, a formation sequence that begins with the thick disc before transitioning to the thin disc, as described in the ``upside-down'' and ``inside-out'' scenario of \citet{Bird:2013aa}, is the predominant theory. In this sequence, disc stars initially form within a thicker and radially compact structure before being born in progressively thinner and larger structures \citep[see, e.g.][]{Bird:2013aa, Wang:2019aa, Buck:2020aa, Bird:2021aa, Agertz:2021aa}.

New observations by the James Webb Space Telescope \citep[JWST;][]{JWST} have provided an unprecedented view into the early Universe, unveiling well-structured disc galaxies at significantly higher redshifts, $3 < z < 8$, than previously thought \citep[e.g.][]{Ferreira:2022aa, Ferreira:2023aa, Kartaltepe:2023aa, Robertson:2023aa, Tohill:2024aa}. Studies with the Atacama Large Millimeter/submillimeter Array \citep[ALMA;][]{ALMA} have also found a large fraction of star-forming dynamically cold discs at as early as $z \sim 6.8$ \citep[e.g.][]{Smit:2018aa, Rizzo:2020aa, Rizzo:2021aa,Roman:2023aa}. Moreover, in recent years, scenarios have been proposed where the thin disc begins to form concurrently with the thick disc, suggesting the presence of a substantial population of old stars in the former \citep[e.g.][and \citealt{Michael:2022aa} for the formation of old cold gaseous discs]{Silva:2021aa, Zhang:2021aa, vanDonkelaar:2022aa}. \citet{Tamfal:2022aa} took it a step further and, using the GigaEris cosmological ``zoom-in'' simulation, showed that thin discs\footnote{A thin disc is defined in \citet{Tamfal:2022aa} as a stellar component made of stars younger than 17~Myr, and whose ratio between the mean rotational velocity, $<v_{\phi}>$, and the radial velocity dispersion, $\sigma_{R}$, is larger than unity. When re-writing the analysis pipeline for this work, we discovered inaccuracies in the calculation of $<v_{\phi}>$ presented in \citet{Tamfal:2022aa}, which led to an overestimate of $<v_{\phi}> / \sigma_{R}$. Since with the revised calculation the values are still  comfortably larger than unity, the main conclusion of \citet{Tamfal:2022aa} is not affected.}, namely kinematically cold discs (KCDs) are already present at redshift $z\sim 7$. This was further supported by the work of \citet{Kohandel:2023aa}, who found dynamically cold gaseous discs using the SERRA \citep{Pallottini:2022aa} suite of zoom-in simulations.  

Despite the  new results, it is evident that the present-day MW thin disc comprises predominantly young and kinematically cold stars \citep[e.g.][]{Parker:2004aa, Wyse:2006aa, Brook:2012aa}. This prompts questions regarding the evolution and fate of the primordial discs detected by ALMA and JWST, as well as those observed  in simulations. The results presented by \citet{vandonkelaar:2023ab} showcase the presence of early KCD stars in the central region of high-redshift disc galaxies, but also suggest that the majority of stars within these primordial discs are unlikely to evolve into the present-day thin disc of spiral galaxies such as our own MW, as both angular momentum transport by a stellar bar and the presence of several massive star clusters that can sink due to dynamical friction underscore the tendency of stars to move inwards. This offers compelling indications that stars within early KCDs experience gradual migration away from their birth locations, a phenomenon known as ``radial migration''.

Interestingly, recent investigations are uncovering the existence of old metal-poor stars, some even reaching ultra metal-poor levels ([Fe/H] $< -4$), with large angular momenta, primarily located close to the Galactic plane and following dynamically cold disc orbits \citep[e.g.][]{Sestito:2019aa, Sestito:2020aa, Sestito:2021aa, Cordoni:2021aa, Dovgal:2024}. Notably, observations with GAIA \citep[by, e.g.][]{Nepal:2024aa} have even come so far to conclude that the MW, similar to the high-redshift galaxies observed by JWST, has an old KCD. The stars in this old disc range from metal-poor to  super metal-rich metallicities, suggesting a history of intense star formation (SF) and the early establishment of cold discs \citep[see also][]{Fernandez:2024aa}.

Recent studies utilizing  JWST have also demonstrated the presence of bulge-dominated galaxies at high redshift \citep[e.g.][]{Huertas:2023aa}, but the main drivers of bulge growth remain largely unknown. \citet{Guedes:2013aa} employed Eris, a ``zoom-in" cosmological simulation of a late-type galaxy, to explore the formation of its pseudo-bulge. They noted that the majority of the pseudo-bulge's mass originated from a bar configuration at early times,  which subsequently transformed into a compact flattened structure and an inner bar. Similarly, \citet{Okamoto:2013aa} examined the formation pathways of two pseudo-bulges within hydrodynamical simulations and showed that both pseudo-bulges formed at high redshift through the accretion of misaligned gas, with secular evolution contributing less than 30 per cent to their final mass. \citet{Gargiulo:2019aa} studied the galactic bulges in the Auriga \citep{Grand:2017aa} simulation and found that pseudo-bulges are formed predominantly in-situ. JWST observations and simulations both seem to suggest that the formation of the primordial discs and bulge occurred within a similar time frame.

In this paper, we examine the fate of primordial KCDs in a MW-sized galaxy. Specifically, we aim to demonstrate that a significant portion of stars born within early primordial thin discs undergo inward migration, resulting in structures with smaller radii, and hinting at the fact that the ``inside-out'' formation scenario of the present-day MW thin disc does not occur at high redshift. Eventually, we argue that these migrated stars integrate into other substructures within the galaxy, particularly the pseudo-bulge. Additionally, we argue that a subset of stars born at the outer edges of the primordial KCDs may persist within the present-day thin disc. Here, we employ an exceptionally high-resolution, hydrodynamical, cosmological ``zoom-in'' simulation \citep[GigaEris;][]{Tamfal:2022aa}. The plan is  as follows: Section~\ref{sec:method} provides a concise overview of the simulation setup. In Section~\ref{sec:result}, we present the results of the simulation, focusing on the trajectory and final location of stars that were originally  ``born within the KCDs''. Finally, we discuss our findings in Section~\ref{sec:discussion}.

\section{Methods}\label{sec:method}

\subsection{The GigaEris simulation}

The $N$-body/hydrodynamic cosmological ``zoom-in'' simulation GigaEris was presented in \citet[][]{Tamfal:2022aa}. The simulation was run using \textsc{ChaNGa} \citep[][]{Jetley:2008aa,Jetley:2010aa, Menon:2015aa}, an $N$-body smoothed-particle hydrodynamics code \citep[][]{Wadsley:2017aa}. The simulation relies on a low-resolution dark matter (DM)-only run, which was used to identify a Galactic-scale halo at $z = 0$. The halo was chosen to closely match the mass of the MW's halo and exhibit a rather quiet late merging history, following a methodology similar to that of the original Eris suite \citep{Guedes:2011aa}. The halo was then re-simulated at higher resolution. The initial conditions were generated with the \textsc{MUSIC} code \citep[][]{Hahn:2011aa}, with 14 levels of refinement and the cosmological parameters $\Omega_{\rm m}$ = 0.3089, $\Omega_{\rm b}$ = 0.0486, $\Omega_{\Lambda}$ = 0.6911, $\sigma_8$ = 0.8159, $n_{\rm s}$ = 0.9667, and $H_0$ = 67.74~km~s$^{-1}$ Mpc$^{-1}$ \citep[see][]{Planck:2016aa}. 

We set the gravitational softening for all particles as a constant in physical coordinates ($\epsilon_{\rm c} = 0.043$ kpc) for redshifts lower than $z = 10$, and scaling as $ 11\epsilon_{\rm c}/(1+z)$ at earlier times. The final snapshot, corresponding to $z=4.4$, contains the following particle counts: $n_{\rm DM} = 5.7 \times 10^8$, $n_{\rm gas} = 5.2 \times 10^8$, and $n_{\star} = 4.4 \times 10^7$. Star particles, each with an initial mass of $m_{\star} = 798$~M$_{\sun}$, were generated based on gas density and temperature criteria \citep[][]{Stinson:2006aa}. The SF rate depends on the gas density, gas dynamical time, and a fixed SF efficiency. The code calculates non-equilibrium abundances and cooling for hydrogen and helium, including gas  self-shielding and a redshift-dependent radiation background \citep{Pontzen:2008aa, Haardt:2012aa}. Additionally, cooling from metallic elements' fine structure lines is computed using rates from Cloudy \citep{Ferland:2010aa, Ferland:2013aa} and the methodology of \citet{Shen:2010aa} and \citet{Shen:2013aa}, assuming no self-shielding \citep[for a discussion, see][]{Capelo:2018aa}.

Feedback from Type Ia and Type II supernovae (SNae) injects energy, mass, and metals into the surroundings \citep[][]{Thielemann:1986, Stinson:2006aa}. The SN Type Ia feedback is independent of progenitor mass, whereas SN Type II feedback follows a delayed-cooling recipe \citep{Stinson:2006aa}. For each SN Type II event, gas, oxygen, and iron mass, dependent on star mass, are injected into the surrounding gas \citep{Woosley:1995aa, raiteri:1996aa}. Stars with masses 8--40~M$_{\sun}$ undergo SN Type II, whereas those with masses 1--8~M$_{\sun}$ release mass as stellar winds, with the metallicity of the wind gas being the same as that of the star particle.

\subsection{Identifying substructures}\label{sec:structures}

\subsubsection{Primordial kinematically cold discs}\label{sec:struc_thin}

\begin{figure}
\centering
\setlength\tabcolsep{2pt}
\includegraphics[ trim={0cm 0cm 0cm 0cm}, clip, width=0.48\textwidth, keepaspectratio]{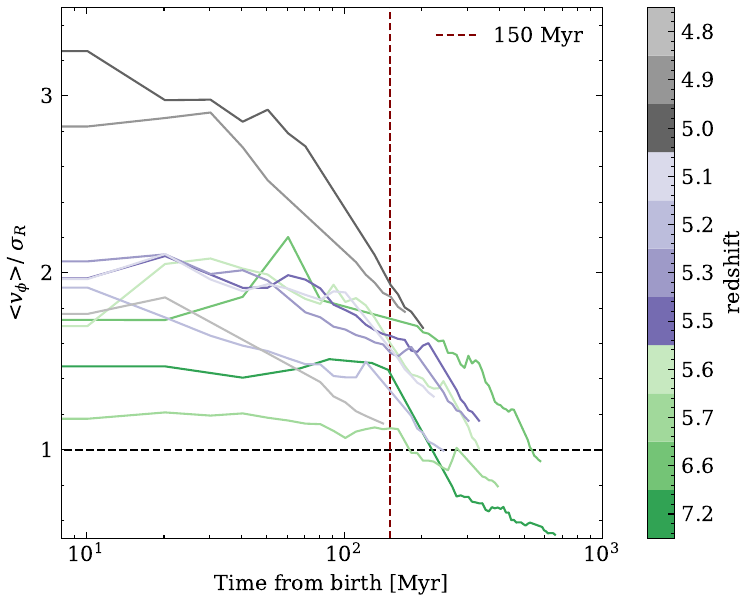}
\caption{Kinematical evolution of the 11 distinct stellar KCDs that formed at various times and along misaligned accretion planes. We show their $<v_{\phi}>$~$/$~$ \sigma_{R}$ values as a function of time. The horizontal black dashed line indicates $<v_{\phi}>$~$/$~$ \sigma_{R} = 1$ and the vertical red dashed line shows our time constraint of 150~Myr. The colour of the line corresponds to the redshift at which the KCD was formed. }
\label{fig:voversigma}
\end{figure}

We examine the evolution of the KCDs by initially decomposing the galaxy into its components at different redshifts. In the study by \citet{Tamfal:2022aa}, multiple stellar discs were identified within the GigaEris simulation at $z>4.4$, emerging at different times and along different, misaligned accretion planes. The discs were identified using the DBSCAN clustering algorithm \citep{Ester:1996aa} for stars born within the last 17~Myr within a 4~kpc box surrounding the galaxy's centre.\footnote{For a more detailed description of the method, see \citet{Tamfal:2022aa}.} In this work, we simply define KCDs as those cold stellar components having $<v_{\phi}>$~$/$~$ \sigma_{R}$ greater than unity and persisting for at least 150~Myr from the moment their first stars are born. We identify 11 distinct stellar KCDs that formed at various times and along misaligned accretion planes (see Figure~\ref{fig:voversigma}), the oldest of such discs originating at $z \sim 7.2$. 

It is worth highlighting the compact radial extent of such discs, spanning no more than 1~kpc (see Section~\ref{sec:link_pseudo-bulge}) at these early epochs. The compact size reflects the small virial radius of the main host halo ($\sim$$37$~kpc at $z = 4.4$) which, in turn, imposes a limit on the orbital angular momentum of baryonic matter that can gravitationally bind to the halo and subsequently cool to form a disc \citep[][]{White:1978aa, Efstathiou:1980aa}. Consequently, the overall aspect ratio (scale height to disc scale length) of the identified stellar KCDs is larger compared to their present-day counterparts, exceeding for example the aspect ratio of the present-day MW thin disc, which is estimated to be between $0.1$ and $0.25$ \citep[e.g.][]{Bland:2016aa}. We also observe a decreasing trend in the aspect ratio of these structures. For instance, the primordial KCD that formed at $z \sim 7.2$ exhibits an initial aspect ratio of $0.49$, whereas the disc formed at $z \sim 4.9$ has an initial aspect ratio of $0.32$. This trend is accompanied by a shift in kinematics, as depicted in Figure~\ref{fig:voversigma}, which shows that the later discs have a higher $<v_{\phi}>$~$/$~$ \sigma_{R}$ ratio when born.

\subsubsection{Pseudo-bulge}

\begin{figure}
\centering
\setlength\tabcolsep{2pt}
\includegraphics[ trim={0cm 0cm 0cm 0cm}, clip, width=0.48\textwidth, keepaspectratio]{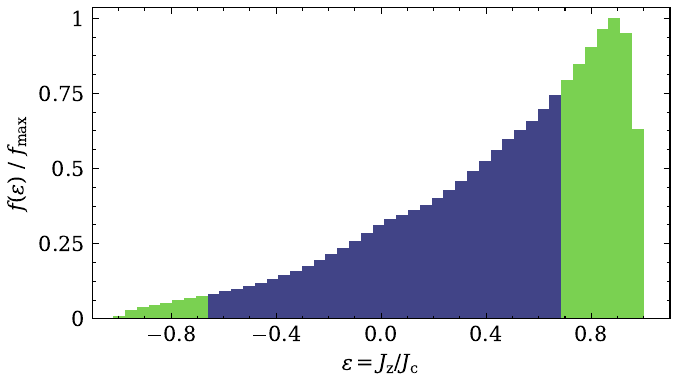}
\caption{Distribution of the circularity parameter $\epsilon$ for stellar particles in the main galaxy halo at the final $z = 4.4$ snapshot, showed as a mass-weighted histogram, which is then normalized to the maximum value of the distribution, $f_{\rm max}$. The purple bins highlight the circularity constraint ($|\epsilon| \leq 0.7$) used to assign stellar particles to the bulge.}
\label{fig:circ}
\end{figure}

We use a combination of spatial and kinematical criteria, similar to those employed by \citet{Soko:2017aa} and \citet{Gargiulo:2019aa}, to define the pseudo-bulge at the final snapshot of the simulation, $z=4.4$.

Firstly, we consider particles located inside a sphere of radius $r_{\rm bulge} = 2 R_{\rm eff}$, where $R_{\rm eff}$ is the effective radius of the bulge (enclosing half of the total projected light in a given band), determined here by fitting a \citet{Sersic_1963} profile to the face-on surface brightness in the $V$-band. We employ a non-linear least-squares approach to fit the sum of an exponential and a \citeauthor{Sersic_1963} function, expressed in terms of projected radiation intensity profile, as

\begin{equation}
    I(r) = I_{\rm e} \exp \Bigl( -b_{\rm n} [(r/R_{\rm eff})^{1/n} -1 ] \Bigr) + I_{\rm 0} \exp \Bigl( - (r/h) \Bigr),
\end{equation}

\noindent where $I_{\rm e}$ is the intensity of the bulge at $R_{\rm eff}$, $n$ is the \citeauthor{Sersic_1963} index, and $b_{\rm n}$ satisfies the equation $\Gamma(2 \rm{n}) = 2 \gamma (2 \rm{n}, b_{\rm n})$, where $\Gamma$ and $\gamma$ are the Gamma function and lower incomplete Gamma function, respectively (to ensure that $R_{\rm eff}$ contains half the projected light). Additionally, $I_{\rm 0}$ stands for the central intensity of the disc component, and $h$ denotes the disc scale length with value $h=1.12$~kpc. We find a \citeauthor{Sersic_1963} index of 0.7 (and $b_{\rm n} = 1.08$). This classifies our bulge as a pseudo-bulge, since pseudo-bulges typically have indices below 2, whereas classical bulges have indices above 2, with minimal overlap between the two categories \citep[see, e.g.][]{Fisher:2008aa}.

Secondly, we use a circularity parameter, $\epsilon$, derived from the $z$-component of the angular momentum vector ($J_z$) of stellar particles within the galactic disc in the $x$-$y$ plane, compared to the angular momentum ($J_{\rm c}$) of a hypothetical particle on a circular orbit: $\epsilon = J_z/J_{\rm c}$. In a typical spiral galaxy, the distribution of circularities -- $f(\epsilon) = \Delta N/\Delta \epsilon$, where $\Delta N$ is the number of particles in the circularity bin $\Delta \epsilon$ around $\epsilon$ -- is expected to exhibit two prominent peaks, one around $\epsilon \approx 1$, corresponding to the disc, and the other at $\epsilon \approx 0$, for the bulge. To define the pseudo-bulge, we exclude stellar particles with pure disc kinematics as well as those in the outer halo. For that purpose, we consider only particles with circularities $|\epsilon| \leq 0.7$, confined within a central cylinder of radius 2~kpc and height 0.5~kpc. The circularity of the stars in the main galaxy halo at the final snapshot is plotted in Figure~\ref{fig:circ}.

\section{Results}\label{sec:result}

\begin{figure}
\centering
\setlength\tabcolsep{2pt}
\includegraphics[ trim={0cm 0cm 0cm 0cm}, clip, width=0.4\textwidth, keepaspectratio]{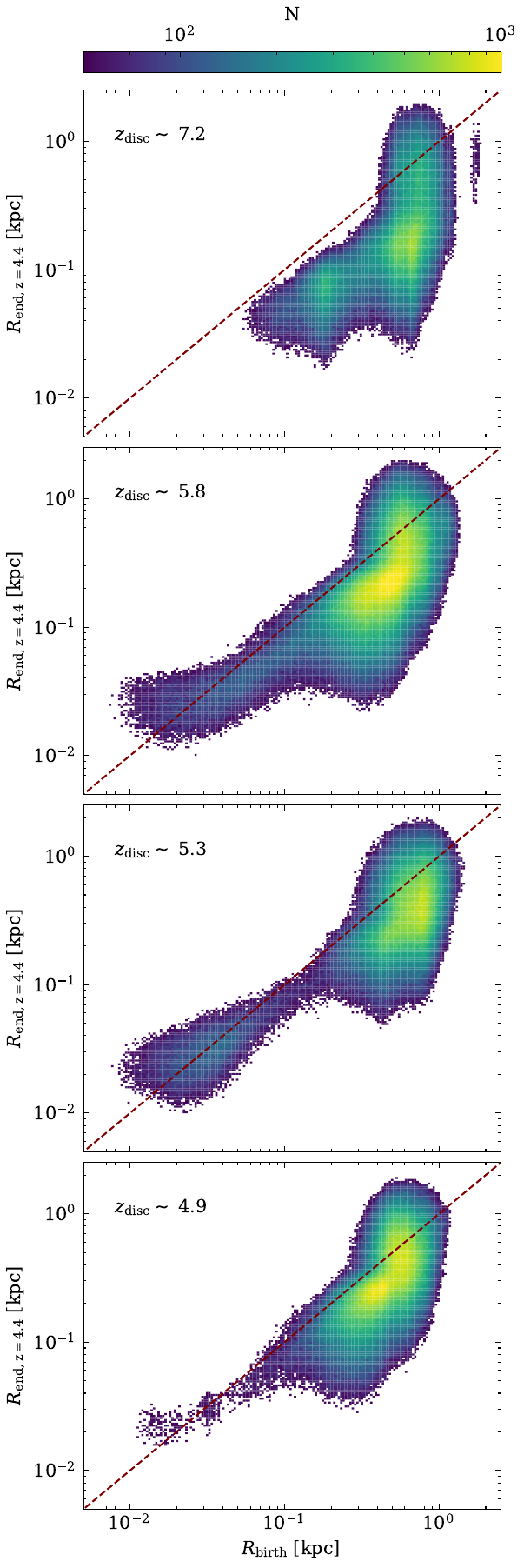}
\caption{The radial migration of KCD stars born at different redshifts is depicted by plotting their birth radius against the radius at the final, $z = 4.4$ snapshot. Stars above the red dashed line migrated outwards, while those below migrated inwards.}
\label{fig:Rmig}
\end{figure}

To determine the fate of the primordial KCDs, we start by looking into the distances from the galactic centre\footnote{Determined by computing the centre of mass at the relevant redshift of all matter within the virial radius.} of  stars both at birth and in the final snapshot. Specifically, we focus on the discs formed at redshifts $z \sim 7.2$, $5.8$, $5.3$, and $4.9$ and plot in Figure~\ref{fig:Rmig} the two distances for all the stars born in such structures. The red dashed lines mark the locus where the birth and final radii are equal, separating outward migration for stars above the line and inward migration for those below it. We utilize 150~bins (evenly sized in logarithmic space) for both radii, spanning from 0.001~kpc to 5~kpc, and implement a strict threshold of at least 50 star particles per bin to make sure we recover general trends and exclude outliers.

\begin{figure}
\centering
\setlength\tabcolsep{2pt}
\includegraphics[ trim={0cm 0cm 0cm 0cm}, clip, width=0.48\textwidth, keepaspectratio]{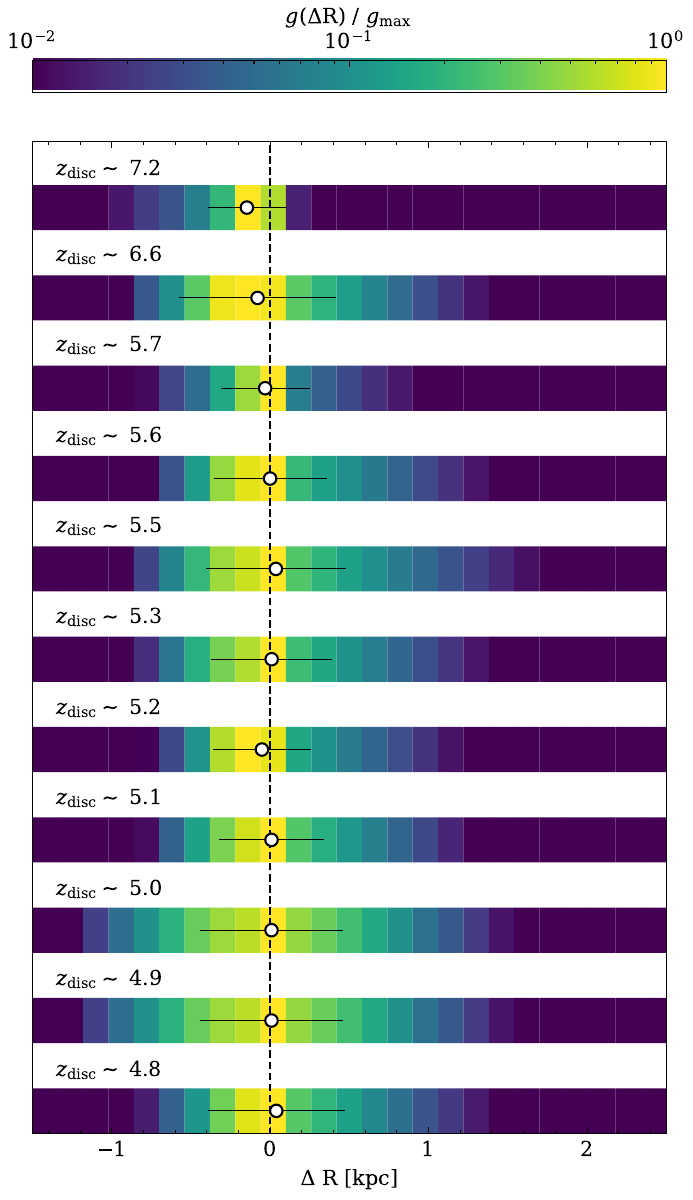}
\caption{We depict the mean disparity (white circle), alongside the $1 \sigma$ standard deviation, between the birth radius and the radius at $z=4.4$ of the KCD-born stars across 11 different identified primordial discs. Furthermore, a mass-weighted histogram is presented, normalized for each separate disc to the distribution's maximum value, $g_{\rm max}$.}
\label{fig:Rdiff_mig}
\end{figure}

Intriguingly, KCD stars born at $z \sim 7.2$ exhibit predominantly inward migration. This pattern starts shifting at $z \lesssim 5.8$, when KCD stars begin displaying a more typical migration activity, involving both inward and outward displacements. This trend is visible in all 11 identified discs (see Figure~\ref{fig:Rdiff_mig}). The peak in concentration of star particles moves closer to the red dashed line for disc stars born at $z \sim 5.3$ and $4.9$. Stellar radial migrations are likely driven by interactions with non-axisymmetric features of the galaxy, such as the bar that is present in GigaEris \citep[as discussed in][]{Tamfal:2022aa, vandonkelaar:2023ab}. These features may induce `churning' \citep[][see also \citealt{Willet:2023aa} for its connection to the thin disc]{Sellwood:2002aa}, wherein an exchange of angular momentum redirects  stars into new orbits. Additional research is however necessary to ascertain the primary driver behind the radial migration of stars in early KCDs.

\subsection{Link to the formation of the pseudo-bulge}\label{sec:link_pseudo-bulge}

\begin{figure}
\centering
\setlength\tabcolsep{2pt}
\includegraphics[ trim={0cm 0cm 0cm 0cm}, clip, width=0.48\textwidth, keepaspectratio]{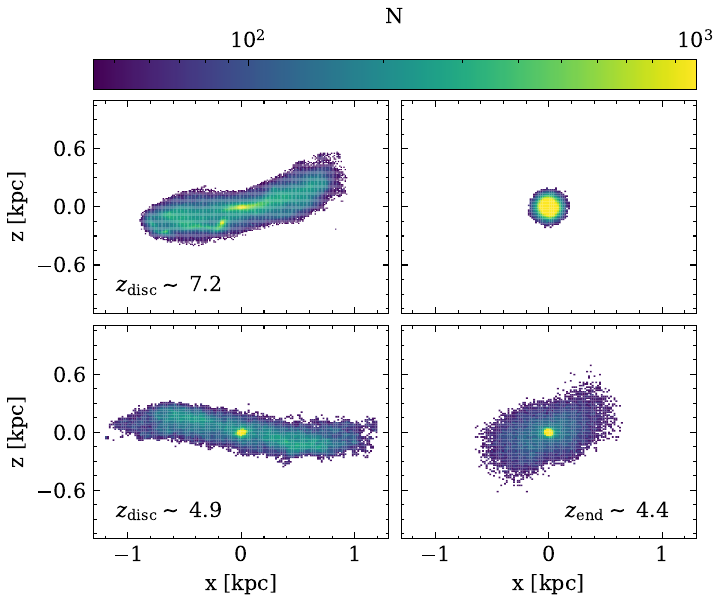}
\caption{Edge-on view of the stars born in a KCD at $z \sim 7.2$ (top-left panel) and of those born in another KCD at $z \sim~4.9$ (bottom-left panel), alongside the same stars at the concluding snapshot of the simulation, at $z = 4.4$ (right-hand panels). We employed 150~bins in both axes, covering a range from -1.2~kpc to 1.2~kpc. Only bins containing a minimum of 50 particles are displayed, highlighting the trend while minimizing noise of the outliers.}
\label{fig:disc}
\end{figure}

We can delve deeper into the information presented in Figure~\ref{fig:Rmig} by examining the distribution of stars within the $x$-$z$ plane\footnote{The $z$-axis is aligned with the angular momentum vector at the specific redshift, computed using all particles within a spherical region whose radius extends to the farthest identified star particle belonging to the primordial KCD from the centre of the galaxy.} of primordial KCDs at birth and at the final snapshot of the simulation at $z = 4.4$, as depicted in Figure~\ref{fig:disc} for the discs born at $z \sim 7.2$ (top panels) and $z \sim 4.9$ (bottom panels). Combining the motion of stars shown in Figure~\ref{fig:Rmig} with Figure~\ref{fig:disc}, we provide the first compelling hints that primordial KCDs at $z \gtrsim 6$ become part of the central region of the galaxy, whereas later discs contribute more to the thick disc. The upper-right panel of Figure~\ref{fig:disc} displays the position of stars originating from a KCD born at $z \sim 7.2$ in the final snapshot of the simulation ($z=4.4$). This panel clearly reveals a clustering of stars at the galaxy's centre, with the majority congregating within a radius of 250~pc. In contrast, stars stemming from a KCD born around $z \sim 4.9$ maintain a more disc-like arrangement, though with an increased thickness and shorter scale radius, as shown in the bottom-right panel. This suggests that the primordial KCDs formed at $z \gtrsim 6$ may play a notable role in shaping central structures like the pseudo-bulge.

\begin{figure}
\centering
\setlength\tabcolsep{2pt}
\includegraphics[ trim={0cm 0cm 0cm 0cm}, clip, width=0.48\textwidth, keepaspectratio]{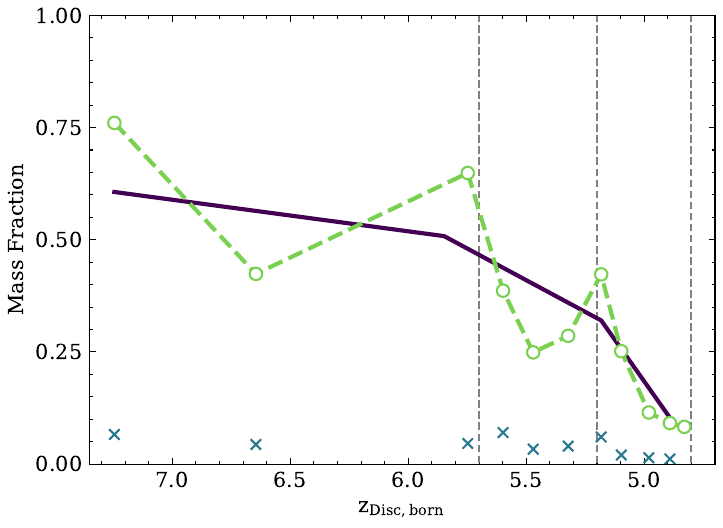}
\caption{The fraction of mass from the ``born in KCDs'' stars contributing to the pseudo-bulge at redshift $z = 4.4$ (green dashed line), whereas the purple line illustrates the overall trend. The blue crosses indicate the fraction of the total mass of the pseudo-bulge identified at $z=4.4$ originated from each primordial KCD. Additionally, the three gray dashed vertical lines mark peaks in SF at $ z \leq 6$ as identified by \citet{Tamfal:2022aa}.}
\label{fig:bulge}
\end{figure}

However, the primordial stellar KCD at $z \sim 7.2$ has a relatively low mass ($\approx 10^{8.79}$~M$_{\sun}$). Therefore, even though 76 per cent of the stellar mass of this structure ($\approx 10^{8.67}$~M$_{\sun}$) ends up in the pseudo-bulge, only about 7 per cent of the total mass of the pseudo-bulge identified at $z=4.4$ ($\approx 10^{9.85}$~M$_{\sun}$) originates from this disc. For all identified primordial KCDs, this mass fraction ranges between 2 and 8 per cent, as shown by the blue crosses in Figure~\ref{fig:bulge}. Consequently, approximately 40 per cent of the stellar mass of the identified pseudo-bulge at $z=4.4$ comes from stars originating from all primordial KCDs that meet our criteria, and all discs contribute more or less the same amount of stellar mass to the central structure. Still, the early primordial KCDs at $z \gtrsim 6$ have a significant impact on the early stages of pseudo-bulge formation. Specifically, 38 per cent of the stars that are already formed at $z \sim 7.2$ and will contribute to the pseudo-bulge originated from the first primordial KCD.

To further investigate this evolutionary scenario, we compute the proportion of the initial KCDs' stellar mass that will ultimately reside in the pseudo-bulge at different formation redshifts using the identified star particles associated with the pseudo-bulge, as depicted in Figure~\ref{fig:bulge} (green, dashed line). This figure shows us indeed that most of the stars born in the primordial KCDs at $z \gtrsim 6$ become part of the pseudo-bulge. It presents a particularly intriguing result, indicating that the vast majority (approximately 76 per cent) of the stellar mass ($\approx 10^{8.67}$~M$_{\sun}$) in the stellar KCD formed at $z \sim 7.2$ will transition to the pseudo-bulge by $z=4.4$, whereas this proportion decreases to approximately 8 per cent for the stellar mass ($\approx 10^{7.68}$~M$_{\sun}$) in the KCD formed at $z \sim 4.8$.

It is evident from Figure~\ref{fig:bulge} that the mass fraction is not a monotonic function of time, as indicated by the peaks around $z \sim 5.7$ and $\sim 5.2$. At these redshifts, a significant portion of the stellar mass formed in the primordial KCDs transitions to the pseudo-bulge by $z = 4.4$. Notably, these peaks closely align with two peaks in SF within the main galaxy halo between $5 \lesssim z \lesssim 6$ \citep[see the SF rate in figure~3 of][]{Tamfal:2022aa}. One plausible explanation is that stars born in primordial KCDs formed near peaks of SF experience greater radial migration as a result of the formation of the galactic bar. The correlation between the bar and central SF has been illustrated, for instance, by \citeauthor{Fanali:2015aa}~(\citeyear{Fanali:2015aa}; see also, e.g. \citealt{Martinet:1997aa, Sakamoto:1999aa, Sheth:2005aa, Ellison:2011aa, Athanassoula:2013aa, Zhou:2015aa, Yu:2022aa, Marioni:2022aa, Zee:2023aa}), who demonstrated that as the bar evolves, gas is funneled towards the central regions of the galaxy, thereby enhancing the central SF rate. A higher SF rate could therefore suggest a more pronounced bar and as a non-axisymmetric feature of the galaxy, the bar likely plays a role in radial migration through its interactions \citep[e.g.][]{Sellwood:2002aa}. However, further investigation is needed to gain a better understanding of this phenomenon, especially as the SF peak around $z \sim 4.8$ appears to lack a similar impact as the previous SF peaks. 

As discussed, the stellar mass of the identified pseudo-bulge at $z=4.4$ comprises out of approximately 40 per cent of stars originating from all primordial KCDs that meet our criteria. Amongst the stars not originating from these discs, the vast majority, around 96 per cent, formed within the pseudo-bulge region, defined as the region within twice the effective radius ($2 R_{\rm eff}$), from gas brought in by the bar within the past 500~Myr. Therefore, one could conclude that the pseudo-bulge in the main galaxy halo of GigaEris is mainly formed in-situ, indicating that nearly all stellar particles were born within the virial radius of the main galaxy halo, whether in a KCD or in the pseudo-bulge region itself, aligning with findings from (pseudo-)bulges within the Auriga simulation \citep[see][]{Gargiulo:2019aa}.

\subsection{Link to the present day's discs}

\begin{figure*}
\centering
\setlength\tabcolsep{2pt}
\includegraphics[ trim={0cm 0cm 0cm 0cm}, clip, width=0.99\textwidth, keepaspectratio]{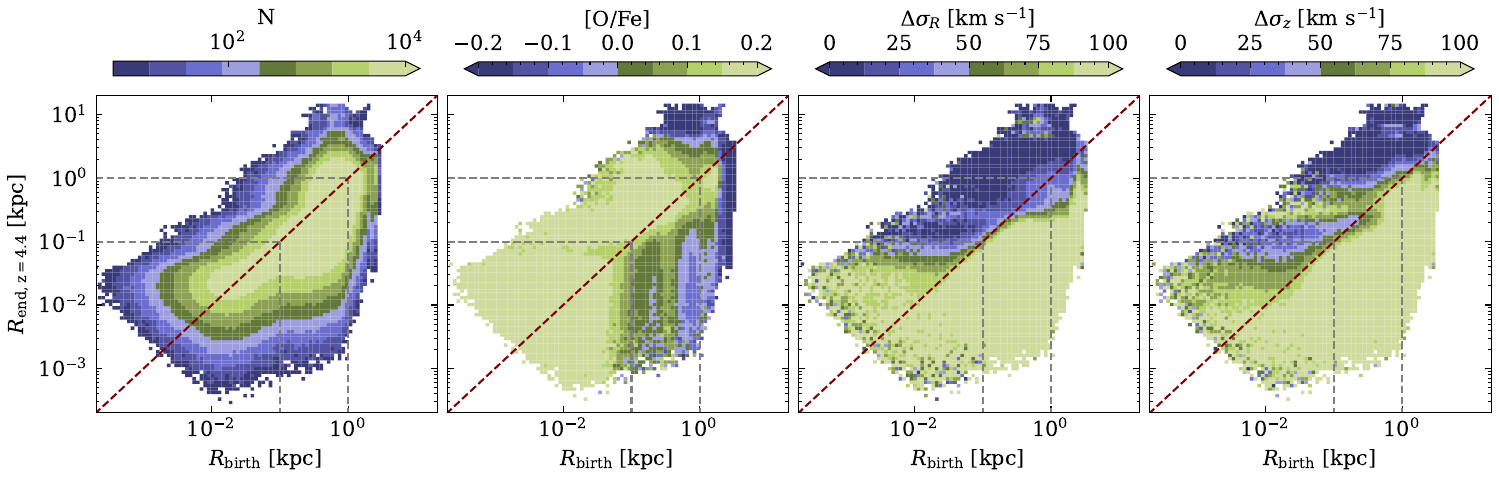}
\caption{The birth radius against the final radius at $z = 4.4$ for all stars born in primordial KCDs, alongside their migration properties. From left to right, the first panel is a 2D histogram for radial migration of KCD stars born at various redshifts, depicted by plotting the birth radius against the radius at the final snapshot at $ z= 4.4$. The second panel illustrates the mean [O/Fe] for the stars within each bin. The last two panels depict the kinematical heating (third panel: radial velocity dispersion; fourth panel: vertical velocity dispersion) of the stars within each bin. The red dashed lines signify where the birth and final radii are equal, indicating that stars above this line migrated outwards, while those below migrated inwards. The gray lines indicate the 0.1~kpc and 1~kpc radii. We utilized 100 bins (equally spaced in logarithmic space) in both axes, covering a range from 0.2~pc to 20~kpc.}
\label{fig:heating}
\end{figure*}

It is important to note that Figures~\ref{fig:Rmig} and \ref{fig:disc} only illustrate the overall trend, as we have not depicted any outliers. The outward migration of stars born within a primordial KCD is common across all redshifts and birth radii, which contrasts with what is shown, for example, in the top panel of Figure~\ref{fig:Rmig}. Given that these stars likely experience less kinematical heating at the outer edges of the galaxy, they may still influence the formation of the older component of the thin disc observed today, as e.g. the stars discussed by \cite{Nepal:2024aa}.

Figure~\ref{fig:heating} illustrates the birth radius plotted against the final radius at $z=4.4$ for all stars born in primordial KCDs, with the colour bar indicating the properties of these stars. The red dashed lines signify where the birth and final radii are equal, indicating outward migration for stars above the line and inward migration for those below it. The first panel is similar to  Figure~\ref{fig:Rmig}, yet in this instance, it shows all stars born in all primordial KCDs. This panel indicates that outward migration happens across all birth radii. However, it is notable that the peak of this plot lies below the red dashed line, suggesting that most stars born within the primordial KCDs tend to migrate inwards.

The second panel shows the [O/Fe] ratio\footnote{In this work, we compute the abundance ratios (e.g. [O/Fe]) normalising them to the solar values provided by \citet{Asplund:2009aa}.} of the stars born in the primordial KCDs. In the MW at $z = 0$, thick-disc stars have a larger oxygen abundance than thin-disc stars with the same [Fe/H], as shown by, e.g. \citet{Franchini:2021aa}. In this panel, we observe that the [O/Fe] ratio decreases with a higher R$_{\rm birth}$. Stars with the lowest [O/Fe] ratios were born outside 1~kpc and then migrated inwards or ended up outside 1~kpc as a result of outwards migration by $z=4.4$.  In general, a low [O/Fe] ratio indicates that the galaxy has undergone an extended or delayed SF history \citep[e.g.][]{Gallazzi:2005aa}. Consequently, it makes sense that the outer regions of the galaxy exhibit a low [O/Fe] ratio, as these areas typically experience more prolonged and less intense SF compared to the central regions. These low-[O/Fe] stars are in line with the stars observed by \citet{Nepal:2024aa}, indicating that these potentially older thin-disc stars exhibit a slightly lower [$\alpha$/Fe], akin to [O/Fe], enhancement compared to the thick-disc stars. Previous observations in small, high-resolution spectroscopic samples have also identified old stars with low [$\alpha$/Fe] ratios \citep[see, e.g.][]{Anders:2018aa, Miglio:2021aa, Gent:2024aa}.

The last two panels show the change in vertical and radial velocity dispersions ($\Delta \sigma = \sigma_{\rm end} - \sigma_{\rm start}$). As previously discussed, it is expected that stars in the thin disc have smaller velocity dispersions compared to those in the thick disc \citep[e.g.][]{Gilmore:2002aa, Soubiran:2003aa, Parker:2004aa, Wyse:2006aa, Mackereth:2019aa}. Therefore, for stars born in the primordial KCDs to remain as thin-disc stars, they should experience minimal kinematical heating. In both panels, we can cleary see that stars reaching the outer radius of 1~kpc, particularly those that migrated there, experience less heating. The minimal kinematical heating observed at the outer edges of our galaxy could be explained by the reduced turbulence experienced by stars in these regions, such as the declining effect of spiral arm and bar heating with radius \citep[see, e.g.][]{grand:2016aa, Mackereth:2019aa}.

This suggest that stars with a low [O/Fe] ratio and minimal kinematical heating could be associated with the formation of the older component of the present-day thin disc, possibly representing the recently identified old stars with low [$\alpha$/Fe] ratios. The presence of a kinematically cold component at the outer edge of the disc suggests that mergers with companions may not have been as effective at kinematically heating the stars \citep[see, e.g.][]{GalandeAnta:2023aa,GalandeAnta:2023ab}, given that several mergers occur -- at $z=6.52$, $6.01$, and $4.53$,  with corresponding mass ratios $0.16$, $0.14$, and $0.12$, respectively \citep{Tamfal:2022aa} -- namely after the formation of the first primordial KCD at $z \sim 7.2$. However, we cannot definitively rule out heating via mergers beyond $z<4.4$. Nonetheless, the data from \citet{Nepal:2024aa}, who highlight the presence of old thin-disc stars, suggests that heating through mergers may indeed be minimal. In a galaxy characterized by a more turbulent history of mergers, these old thin-disc stars might not always persist till late times. However, our MW is believed to have experienced a relatively quiescent merger history \citep[e.g.][]{Fragkoudi:2020aa, Kruijssen:2020aa}, making our analysis relevant.

Furthermore, as mentioned earlier, it is important to highlight that the primordial KCDs in our simulations have a characteristic radius of about 1~kpc. Considering that radial migration is expected to still occur beyond $z \lesssim 4.4$, it adds uncertainty about where these surviving KCD stars will ultimately settle. It is reasonable to assume that, to preserve their cold kinematics, they would tend to end up near the present-day outer edge of the thin disc, where they are less affected by secular heating due to the galactic bar and spiral structure. Alternatively, as the current GAIA observations of \citet{Nepal:2024aa} suggest, they might settle around 1~kpc from the galactic center, within the bulge region, where they would experience  little secular heating as this region is dominated by the spheroidal bulge potential, which suppresses the propagation of density waves and weakens the effect of the bar, whose major axis is much larger in scale (5--6~kpc).

\section{Discussion and conclusions}\label{sec:discussion}

In this study, we have demonstrated that most stars within primordial KCDs, characterized as stellar discs with $<v_{\phi}> / \sigma_{R}$ larger than unity for at least 150~Myr after birth, undergo migration and ultimately integrate into other substructures within the galaxy, with particular emphasis on the pseudo-bulge. Stars born in KCDs formed at $z \gtrsim 6$ predominantly migrate to the pseudo-bulge. In contrast, stars born in KCDs formed after this redshift show signs that they might contribute more significantly to the thick disc. However, due to the limitations of the simulations, the fate of these stars at $z \leq 4.4$ remains uncertain. Nevertheless, stars originating from the outer edges of the discs, particularly those that have migrated even farther, exhibit minimal kinematical heating, suggesting a possibility of their survival and integration into the present-day thin disc.

Upon initial examination, the formation mechanism described in this study appears to deviate from the ``inside-out'' and ``upside-down'' formation processes outlined by  \citeauthor{Bird:2013aa} (\citeyear{Bird:2013aa}; see also, e.g. \citealt{Agertz:2021aa, Bird:2021aa, Belokurov:2022aa}) using the Eris simulation. As concluded by \citet{Tamfal:2022aa}, the GigaEris simulation does not exhibit an ``upside-down'' disc formation. Additionally, we demonstrate in this paper that most of the stars (see Figure~\ref{fig:Rdiff_mig}) formed within an early KCD undergo inward migration, so the simulation also does not exhibit an ``inside-out'' disc formation. Nevertheless, while a KCD with vertical extent below 0.5~kpc is present since $z \gtrsim 7$, the disc aspect ratio is larger than in a typical present-day disc of a spiral galaxy, as its initial radial extent is very compact (about a kpc) due to the lack of high angular momentum material at high redshift. Therefore, we suggest that the majority of stars within these primordial KCDs will not contribute to the formation of the present-day thin disc. Instead, the prevalence of ``upside-down'' and ``inside-out'' thin-disc formation mechanisms likely becomes more pronounced at $z \lesssim 4$, extending beyond the final snapshot of our simulation. We therefore argue for a ``two-phase'' migration scenario, wherein the stars born in the primordial KCDs will mainly migrate inwards, leading to new structures with smaller radii than their formation radius.

This ``two-phase'' thin-disc formation process introduces an interesting contrast between the dynamics characterising the early and low-redshift stages of thin-disc evolution. As a consequence, it becomes important to exercise caution when drawing parallels between high-redshift discs observed with instruments like JWST \citep[see, e.g.][]{Ferreira:2022aa, Ferreira:2023aa, Kartaltepe:2023aa, Robertson:2023aa} and their counterparts at lower redshifts. Likewise, since this first phase of disc formation is fast, resulting in a transition to a pseudo-bulge in less than a billion years, the qualitative implication for high-redshift observations is that one should observe a rapid rise of the bulge-to-disc ratio of discy galaxies already at $z> 4$. The timing of the transition would depend on the assembly history of the galaxy, in particular how rapidly the stellar mass can grow enough to become susceptible to internal instabilities such as bar formation, which promote radial migration and pseudo-bulge formation. Our single ``zoom-in'' simulation should thus be seen as a numerical experiment exposing this new evolutionary scenario, while future work will have to assess its quantitative implications and observational diagnostics in a galaxy population at high redshift, by combining information from additional high-resolution ``zoom-in'' simulations and from cosmological volumes. 

However, it is essential to acknowledge again that Figures~\ref{fig:Rmig} and~\ref{fig:disc} present the general trend, without depicting any outliers. The outward migration of stars originating within a primordial KCD is observed across all redshifts, although it may be less frequent as shown by the first panel of Figure~\ref{fig:heating}. As seen in the third and final panel of Figure~\ref{fig:heating}, it is plausible to conclude that the stars that migrated to the outer edge of the galaxy experienced less kinematical heating. This suggests that these stars might have had an influence on the formation of the older component of the thin disc observed today and could potentially represent the previously detected old stars with low [$\alpha$/Fe] ratios \citep[see, e.g.][]{Anders:2018aa, Miglio:2021aa, Gent:2024aa, Nepal:2024aa}. These stars migrated outwards, aligning with the conventional ``inside-out'' disc formation theory. This simultaneous outward migration, alongside inward migration, provides additional support for the proposed ``two-stage'' disc formation process proposed in this paper.

Finally, further investigation is crucial to understand the shift occurring around $z \lesssim 6$. Before this time, most of the stellar mass within the primordial KCDs migrates to the pseudo-bulge region, whereas afterwards less than half of the total stellar mass from the formed primordial KCDs goes towards the pseudo-bulge. This trend intensifies around $z \lesssim 5$, with only about 15 per cent of the stellar mass of the KCD formed at $z \sim 5.0$ ending up in the pseudo-bulge. One potential explanation for this trend is that at redshift $z \lesssim 6$ the atomic neutral hydrogen gas fraction ($F_{\rm gas, HI}$) in the galaxy reaches approximately 25 per cent. As the redshift decreases further to $z \lesssim 5$,  $F_{\rm gas, HI}$ in the galaxy halo drops below 20 per cent. This specific gas fraction has been associated with the conditions necessary for a disc to form that exhibits kinematics similar to those observed in the present-day thin disc \citep[][]{vanDonkelaar:2022aa}. 

\section*{Acknowledgements}
PRC, LM, and FvD acknowledge support from the Swiss National Science Foundation under the Grant 200020\_207406. Support for this work was provided also by NASA through grant 80NSSC21K027 (PM). The simulations were performed on the Piz Daint supercomputer of the Swiss National Supercomputing Centre (CSCS) under the project id s1014.

\section*{Data Availability}
The data underlying this article will be shared on reasonable request to the corresponding author.



\bibliographystyle{mnras}
\bibliography{output} 





\bsp	
\label{lastpage}
\end{document}